# Phase Switchable Photocatalytic Water Splitting via a Paraelectric-Ferroelectric Transition in $Zr_2Ge_2S_6$ Monolayer: A Comprehensive Theoretical Insights

Jubair Hossan Abir[1], Tauhidur Rahman[1], Md. Tanvir Khan[1], S.S.B. Pallab[2],
Raihana Shams Islam[1*], Saleh Hasan Naqib[1*]

[1]Department of Physics, University of Rajshahi, Rajshahi 6205
[2]Department of Electrical and Electronic Engineering, University of Rajshahi, Rajshahi 6205
*Corresponding authors; Emails: ***salehnaqib@yahoo.com**, ***rsislam@ru.ac.bd**

## Abstract

Photocatalytic water splitting (PWS) is a promising technology for addressing the global energy crisis and producing renewable and clean hydrogen fuel. Although numerous 2D materials have recently been proposed as potential photocatalysts, effective strategies for regulating photocatalytic reactions and improving energy conversion efficiency remain limited due to performance regulation challenges. Here, using first-principles calculations, we demonstrate that the photocatalytic activity and energy conversion efficiency of a $Zr_2Ge_2S_6$ monolayer can be effectively tuned through a paraelectric-ferroelectric phase transition. The $Zr_2Ge_2S_6$ monolayer exhibits excellent structural stability, favorable mechanical properties, a suitable band gap, optimal band edge positions, and broad-spectrum light absorption. Moreover, the $Zr_2Ge_2S_6$ monolayer exhibits a higher oxidation potential and a stronger driving force for photogenerated holes to promote oxygen evolution reaction (OER) in the ferroelectric phase. In contrast, the paraelectric phase provides photogenerated electrons with a greater reduction potential and driving force for hydrogen evolution reaction (HER). The solar-to-hydrogen conversion efficiency is also strongly influenced by the phase transition, increasing from 7.71% in the paraelectric phase to 15.31% in the ferroelectric phase because of the improved carrier utilization. Our theoretical investigation not only highlights the crucial role of ferroelectric polarization in photocatalytic water splitting but also provides an effective strategy for tuning the photocatalytic properties of 2D ferroelectric materials through ferroelectric switching.

## 1 Introduction

Across the world, fossil fuels remain the dominant source of energy. Over the past few decades, growing awareness about the environmental damage and atmospheric pollution caused by fossil fuels has encouraged governments, international organizations, institutions, and industries to work together and invest into the development of alternative energy technologies. The main objective is to lessen reliance on finite energy resources while supporting the sustainable development of human society over the long term [**1**]. Hydrogen is increasingly regarded as a promising clean-energy carrier because of its high energy density, abundant availability, zero carbon dioxide emissions during use, and ability to be converted directly into electricity, thermal, or mechanical power [**2**–**6**]. Hydrogen can be produced through several routes, including fossil-fuel processing, methane cracking, catalytic steam reforming, heavy-oil partial oxidation, gasification, water electrolysis and photocatalytic water splitting (PWS). However, many of these methods are expensive, require complicated setups, or are harmful to the environment [7,8]. The minimum energy required for water oxidation is determined by the standard Gibbs free-energy change of overall water splitting ($\Delta G^{\Theta} \approx 237$ kJ mol$^{-1}$), and this high thermodynamic requirement, together with the complex multi-electrons/protons transfer process, contributes to the low quantum efficiency of PWS [9,10]. Accordingly, considerable research over last decades has focused on developing efficient photocatalysts for water splitting, consisting of three-dimensional (3D) porous materials such as metal-organic frameworks, metal-modified Carbon Nitrides, metal oxides and so on [11–13]. In two-dimensional (2D) semiconductors, the surface-to-volume ratio is maximized that exposes numerous active sites, while dangling bonds at edges and defect sites can further enhance their surface reactivity during photocatalytic reactions [14]. The atomically thin geometry of 2D materials allows photogenerated electrons and holes to migrate along shorter transport pathways to surface reaction sites, which helps to reduce electron-hole recombination, and makes these materials suitable candidates for PWS [15–18].

Despite these advantages, most 2D photocatalysts, photogenerated electrons and holes are confined to the same surface, causing hydrogen evolution reaction (HER) and oxygen evolution reaction (OER) to occur at identical sites, which promotes electron-hole recombination and reduces energy conversion efficiency [19]. Consequently, recent studies on 2D semiconductor photocatalysts have primarily aimed at enhancing electron-hole separation efficiency.

Unlike conventional catalysts, ferroelectric materials provide an intrinsic built-in electric field and polarization-induced surface charges, which promote photogenerated carrier separation and enhance interfacial redox capabilities, and thereby improve overall catalytic performance [20–24].

For instance, Janus group III monochalcogenide $M_2XY$ (M = Ga and In and X/Y = S, Se, and Te) monolayers have emerged as promising candidates for PWS due to their low exciton binding energies and efficient electron-hole (e-h) separation facilitated by the internal electric field [25]. Moreover, switchable ferroelectric polarization enables dynamic modulation of catalytic reaction pathways, thereby offering precise control over production selectivity [26].

Notably, 2D ferroelectric materials are highly advantageous for photocatalysis because of their excellent stability, elevated Curie temperatures, large reaction surface areas, and high carrier mobilities [21,26]. Previous studies have shown that host-atom substitution induces switchable polarization in asymmetric single-layer such as $Sc_2CXY$ [27], $M_4X_3Y_3$ [28], $In_2Se_3$ [29], $CuInP_2S_6$ [30] and GeS [31], thereby regulating charge transfer, modifying catalytic mechanisms, and enabling alternative reaction pathways with tunable catalytic activity.

Moreover, ferroelectric phase transition driven polarization reversal has been theoretically and experimentally demonstrated in 2D ferroelectric systems such as $AgBiP_2Se_6$ [32], $BaVNO_2$ monolayers [33], and $Sc_2CO_2/PtS_2$ heterostructures [34], thereby enhancing photocatalytic performance. These findings highlight polarization switching as a promising strategy to enhance solar-to-hydrogen conversion and enable controllable photocatalysis in asymmetric 2D materials. Nevertheless, controlling photocatalysis activity remains challenging because switchable polarization must preserve favorable reaction energetics. For example, in monolayer $Hf_2Ge_2S_6$ [35] the vertical displacement of Ge atoms breaks inversion symmetry, generates an out of plane electrostatic potential difference and switches the preferred photocatalytic pathway between HER and OER.

Based on these insights, the $Zr_2Ge_2S_6$ monolayer is proposed as a promising candidate for photocatalytic applications because of its appropriate band gap, effective solar-light utilization, and suitable band alignments with respect to the redox potentials required for water splitting. In this work, we systematically investigate and compare the electronic structure and photocatalytic performance of $Zr_2Ge_2S_6$ in both ferroelectric (FE) and paraelectric (PE) phases using density functional theory (DFT) with the HSE06 hybrid functional. The results indicate that the FE phase of $Zr_2Ge_2S_6$ exhibits excellent potential for the water oxidation reaction, whereas the PE phase is more favorable for the hydrogen evolution reaction. More significantly, FE switching can effectively regulate the photo-redox capability of the material. Our work demonstrates that the PE-FE phase transition provides an effective strategy for tuning and enhancing the photocatalytic performance of 2D ferroelectric materials.

## 2 Computational methodology

First-principles calculations were performed using density functional theory (DFT) as implemented in the Vienna Ab-initio Simulation Package (VASP) [36,37]. The electron-ion interactions were described through the projector augmented-wave (PAW) method [38]. Exchange-correlation effects were treated using the Perdew-Burke-Ernzerhof (PBE) functional [39] within the generalized gradient approximation (GGA). The Heyd-Scuseria-Ernzerhof hybrid functional (HSE06) [40,41] was further employed to obtain more accurate electronic band gaps. Under periodic boundary conditions, the supercell contained one unit cell of the hexagonal honeycomb of $Zr_2Ge_2S_6$ monolayer with a vacuum spacing exceeding 20 Å in the vertical direction to avoid interactions between adjacent supercells. The convergence criteria are enhanced by using a tested plane-wave kinetic-energy cutoff of 500 eV and the Brillouin zone sampling resolution of $2\pi \times 0.03$ $Å^{-1}$. Due to the intrinsic polarization, self-consistent dipole layers were introduced at the center of the vacuum region to compensate for the difference in vacuum levels on the two sides of the monolayer supercell. The long-range van der Waals (vdW) interactions between layers were incorporated through the DFT-D3 dispersion correction to the total energy [42]. The convergence tolerance is $10^{-5}$ eV for total energy and 0.005 eV/Å for all forces. The climbing image nudged elastic band (CI-NEB) method [43,44] was used to determine the energy barrier for the most favorable ferroelectric switching pathway. Ab initio molecular dynamics (AIMD) simulations were performed to examine the thermal stability of both paraelectric and ferroelectric $Zr_2Ge_2S_6$ monolayers. The simulations were conducted in the canonical NVT ensemble using a Nosé-Hoover thermostat. A $2 \times 2 \times 1$ supercell was simulated at 300 K for 5.0 ps with a time step of 1.0 fs.

The Quantum ESPRESSO package [45] was used to perform DFPT calculations [46] and obtain the phonon spectrum of monolayer $Zr_2Ge_2S_6$. The interactions between the atomic nuclei and valence electrons were described using optimized norm-conserving Vanderbilt pseudopotentials from the PseudoDojo library [47]. A total energy convergence threshold of $10^{-9}$ Ry was adopted with a $12 \times 12 \times 1$ *k*-point grid and an energy cutoff of 80 Ry. A $2 \times 2 \times 1$ *q*-point grid was employed to calculate the lattice dynamical properties.

# 3 Results and discussion

## 3.1 Geometric configuration, dynamic and thermal stability

**Figure 1** presents the FE and PE configurations of a 2 × 2 × 1 $Zr_2Ge_2S_6$ supercell. In both phases, the Zr atoms form a hexagonal honeycomb lattice while the Ge atoms occupy the centers of the hexagons. The PE phase adopts an ideal centrosymmetric structure with the $P\bar{3}1m$ space group (**No. 162**). The Zr atomic plane perpendicularly bisects each Ge pair positioned on the upper and lower surfaces. A small displacement of the lower Ge atoms along the *z* direction breaks the inversion symmetry and transforms the structure into the FE phase with $P31m$ symmetry (**No. 157**). This subtle structural distortion leads to remarkably different physical properties in the two phases.

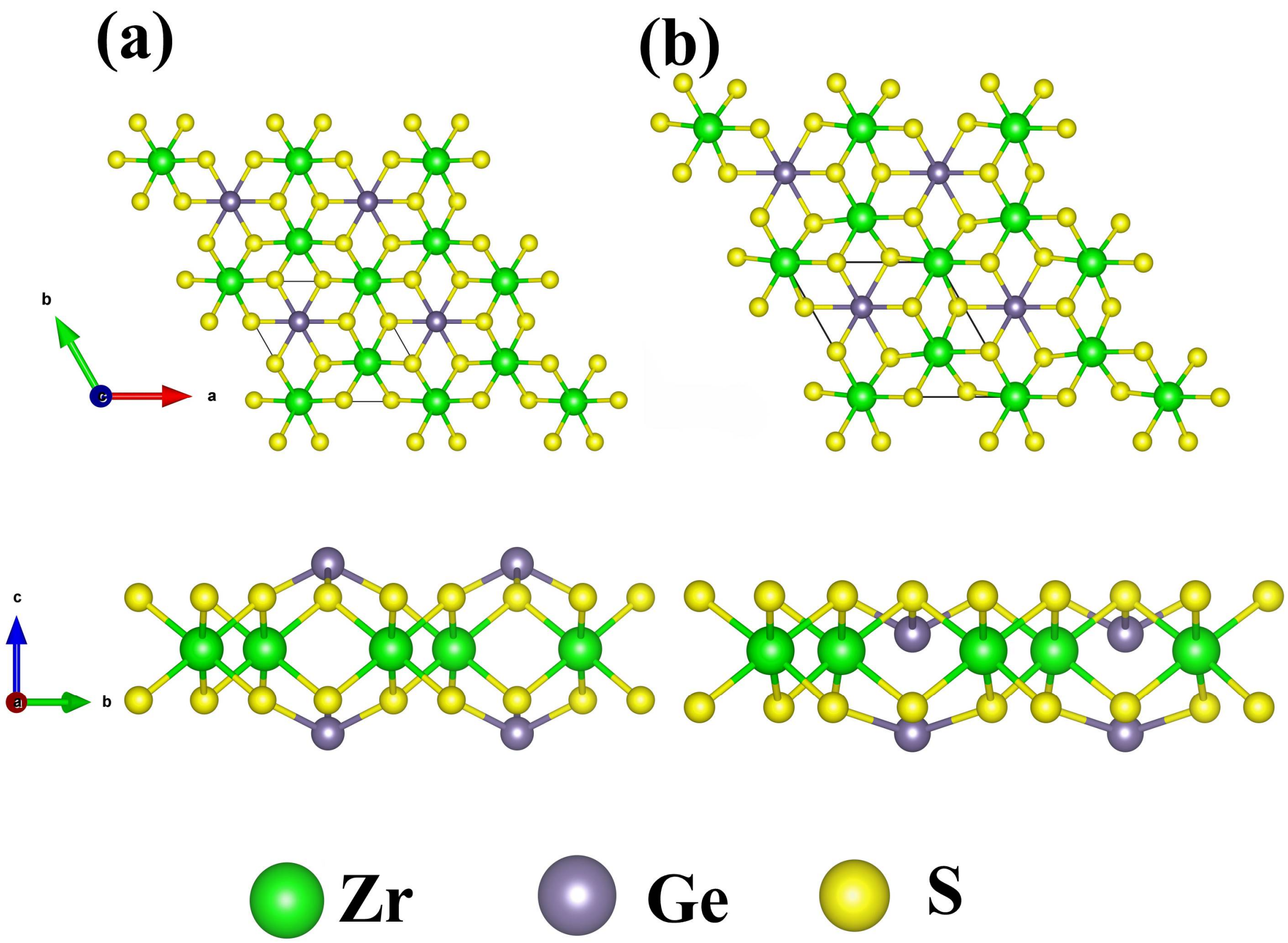

**Figure 1.** The optimized structure of monolayer $Zr_2Ge_2S_6$, where (**a**) the left side is a paraelectric (PE) phase and (**b**) the right side is a downward polarized ferroelectric (FE) phase. The parallelogram indicates two $ZrGeS_3$ primitive cells.

**Table 1.** The lattice parameter (*a*) and bond length (*d*) of paraelectric and ferroelectric $Zr_2Ge_2S_6$ monolayers.

| Phase | *a* (Å) | $d_{Zr-Zr}$ (Å) | $d_{Zr-Ge}$ (Å) | $d_{Zr-S}$ (Å) | $d_{Ge-Ge}$ (Å) | $d_{Ge-S}$ (Å) |
|---|---|---|---|---|---|---|
| $Zr_2Ge_2S_6$ (PE) | 6.38 | 3.68 | 4.42 | 2.56 | 4.89 | 2.41 |
| $Zr_2Ge_2S_6$ (FE) | 6.54 | 3.78 | 3.80/4.37 | 2.55/2.62 | 2.63 | 2.37/2.54 |

**Figure 2** illustrates the electric field driven FE switching pathway and the corresponding energy barrier of 2D $Zr_2Ge_2S_6$, evaluated using the Climbing Image-Nudged Elastic Band (CI NEB) method [**43**,**44**]. The energy profile exhibits two minimum points (0 eV) which correspond to the two equivalent oppositely polarized FE phases. Two transition states appear at 0.89 eV and 0.80 eV along the switching pathway. The maximum switching barrier is approximately 0.89 eV and lies within the ideal range of 0.1 ~ 1.0 eV reported in previous studies [**48**]. The ferroelectric polarization originates from the vertical displacement of the Ge dimer along the *z*-direction [**49**]. This structural feature suppresses unwanted quantum tunneling while allowing polarization reversal under lower applied electric field. The downward polarized FE phase was selected for the subsequent investigation of its electronic, optical, mechanical and photocatalytic properties.

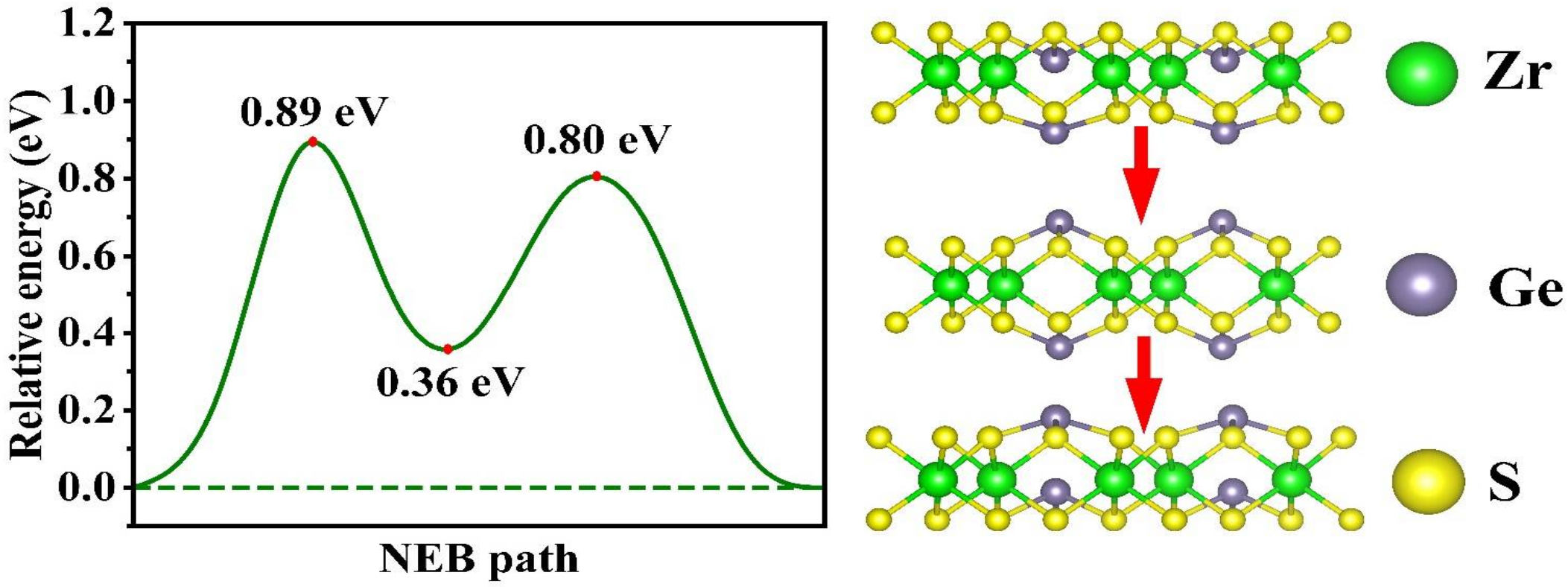

**Figure 2.** The most efficient ferroelectric switching pathway and its corresponding energy barrier. The polarization reverses from the downward polarized FE phase through the intermediate PE phase to the upward polarized FE phase.

The dynamical stability of the paraelectric and ferroelectric phases was examined by calculating their phonon dispersion spectra using density functional perturbation theory (DFPT). As shown in **Figure 3**, no significant imaginary modes are observed throughout the Brillouin zone except for a shallow imaginary pocket near the Γ point in the flexural acoustic branch. Such minor softening is highly sensitive to computational parameters and is commonly encountered in first-principles phonon calculations of two-dimensional materials pointed out by Fal'ko et al. [**50**]. Previous studies have shown that this numerical artifact can be reduced by using a denser *q*-point mesh and stricter convergence settings [**35**]. Therefore, the phonon spectra support the dynamical stability of both phases.

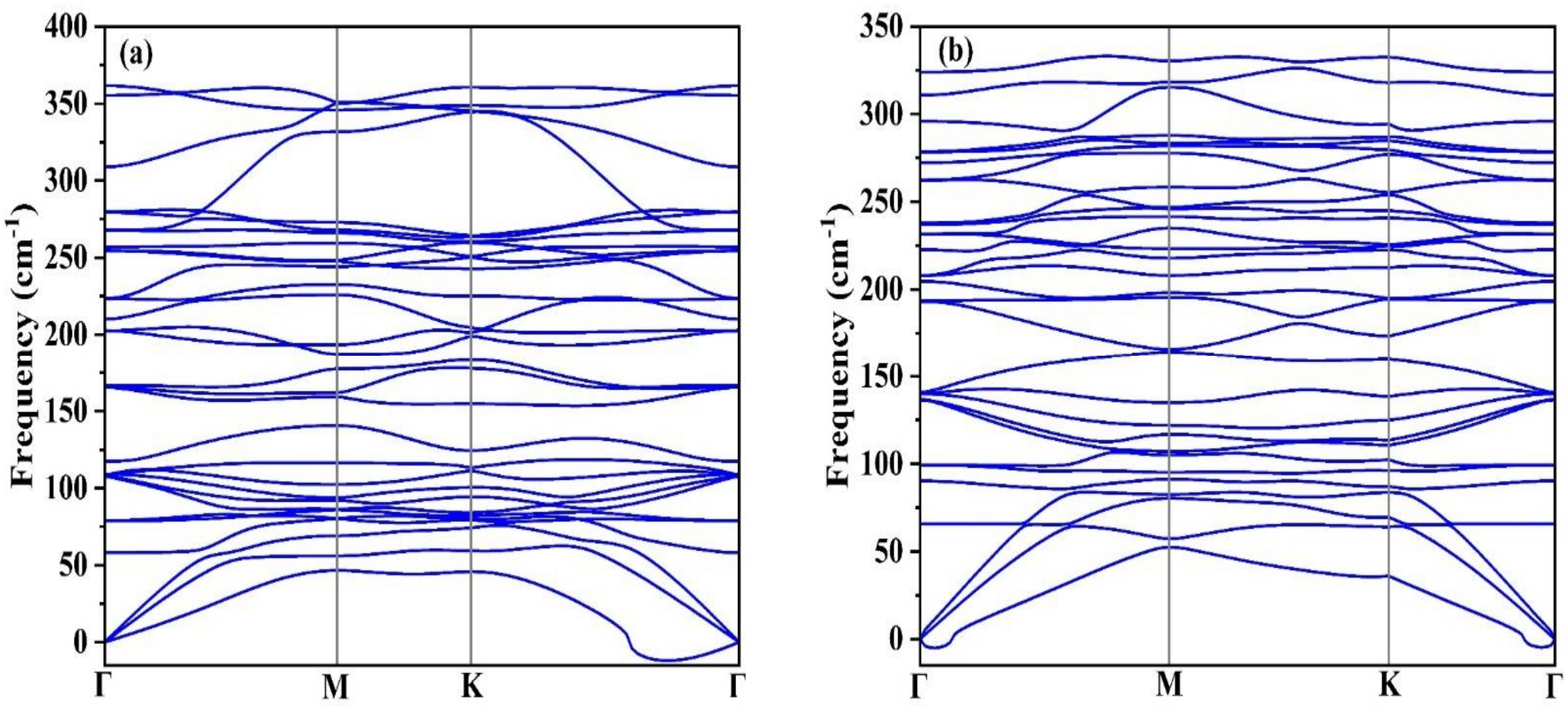

**Figure 3. Dynamic stability.** Phonon dispersion curves of (**a**) PE and (**b**) FE phase of monolayer $Zr_2Ge_2S_6$.

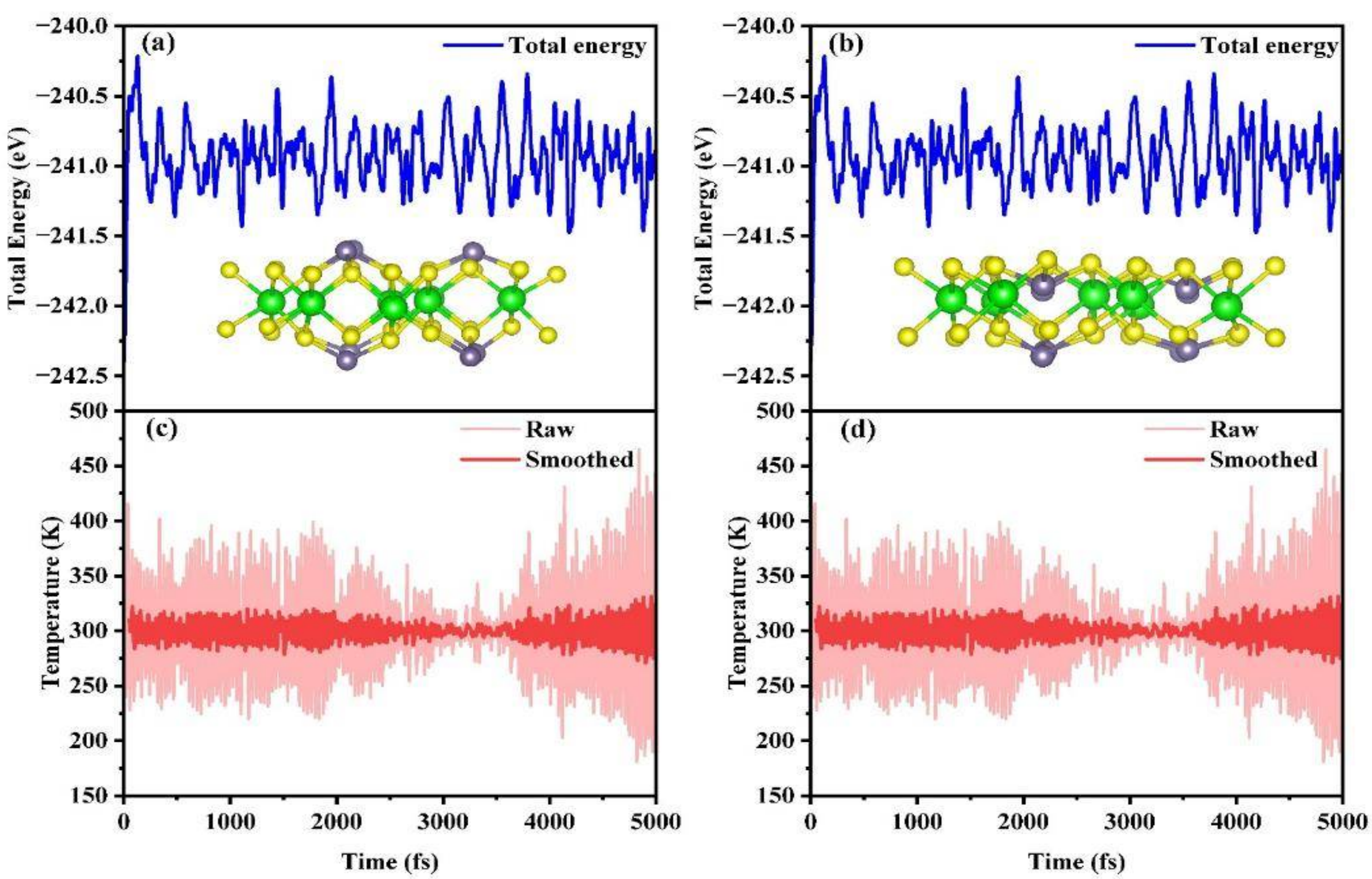

**Figure 4.** Ab initio molecular dynamics simulations of 2D $Zr_2Ge_2S_6$ at 300 K. Time-dependent total energy profiles of (**a**) PE and (**b**) FE phase. Temperature fluctuation profiles of (**c**) PE and (**d**) FE phase, respectively.

The AIMD simulations were performed at 300 K for 5 ps to evaluate the thermal stability of the both PE and FE phases. As shown in **Figure 4**, the total energy fluctuates within a limited range without any noticeable energy drift in both phases. Moreover, the final atomic configurations show no structural distortion or bond breaking. These results indicate that both phases are thermally stable at 300 K.

## 3.2 Electronic properties

The electronic structures of the PE and FE phases of $Zr_2Ge_2S_6$ were investigated using orbital projected band structures and partial density of states, as shown in **Figure 5**. Both phases exhibit indirect semiconducting characteristics. In the PE phase, the valence band maximum is located at the *Γ*-point, whereas the conduction band minimum appears near the *M*-point resulting in a band gap of 1.54 eV. For the FE phase, the valence band maximum lies along the *Γ*-*M* direction, while the conduction band minimum is positioned at the *K*-point producing a larger band gap of 1.81 eV. Thus, PE to FE phase transition increases the band gap by approximately 0.27 eV. The narrower band gap of the PE phase may promote lower energy photoexcitation and stronger visible light absorption. Both band gaps exceed the thermodynamic water splitting energy of 1.23 eV which suggests $Zr_2Ge_2S_6$ holds excellent promise for providing sufficient energy for photocatalytic water oxidation and reduction energy range. The projected density of states (PDOS) reveals similar orbital contributions in both the FE and PE phases of monolayer $Zr_2Ge_2S_6$. The valence band edge mainly arises from hybridized Ge *s* and S *p* states while the conduction band edge is dominated by Zr *d* orbitals with additional contributions from Ge *p* and S *p* states.

**Table 2.** Direct band gap $\boldsymbol{E_g^d}(\mathbf{HSE06})$ and indirect band gap $\boldsymbol{E_g^i}(\mathbf{HSE06})$ and difference of electrostatic potential (ΔΦ) between top and bottom surfaces for PE and FE $Zr_2Ge_2S_6$ monolayers.

| Phase | $E_g^i$(HSE06) (eV) | $E_g^d$(HSE06) (eV) | ΔΦ (eV) |
|---|---|---|---|
| PE | 1.54 | 1.74 | 0.00 |
| FE | 1.81 | 2.01 | 0.58 |

To explore the photoexcitation behavior of both phases, the charge density distributions associated with the VBM and CBM of $Zr_2Ge_2S_6$ were analyzed. The charge density plots show that the PE phase possesses nearly symmetric CBM and VBM distributions on both sides of the monolayer. Such extensive overlap provides little driving force for directional carrier separation. After the transition to the FE phase, the displacement of Ge atoms breaks the equivalence of the two surfaces and induces an asymmetric redistribution of the band-edge states. In the FE phase, the CBM charge density is mainly localized around the Zr-centered regions, with additional contributions from the bottom Ge and top S atoms. In contrast, the VBM charge density is predominantly concentrated around the Ge and S atoms on the bottom surface. Almost no visible VBM charge accumulation occurs around the Zr atoms at the selected isosurface value. This distinct atomic and spatial localization of the CBM and VBM reduces their overlap and may suppress photogenerated electron-hole recombination. Furthermore, the polarization-induced internal electric field can promote the directional migration of electrons and holes toward different reactive regions of the monolayer resulting in efficient charge separation [**51**].

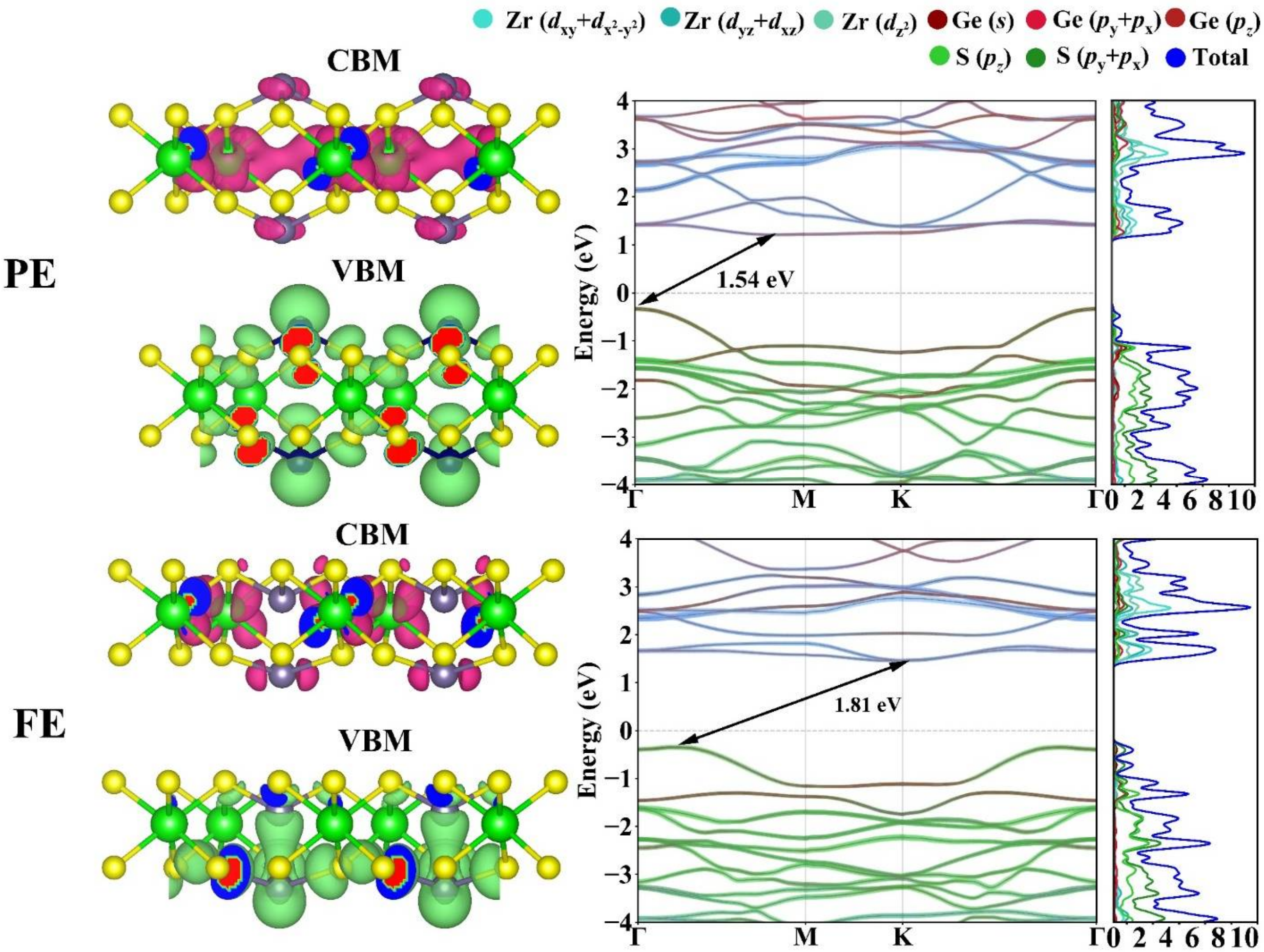

**Figure 5.** The right panels represent the electronic band structures of PE phase and FE phase of monolayer $Zr_2Ge_2S_6$ at HSE06 level, respectively. The left panels represent the orbital distribution at conduction band minimum (CBM) and valence band maximum (VBM) of the corresponding phases.

## 3.3 Optical properties

Strong optical absorption is a fundamental requirement for a photocatalytic monolayer because the production of photoexcited electron-hole pairs depends on how effectively the material interacts with incident photons. Therefore, the optical behavior of the FE and PE phases of monolayer $Zr_2Ge_2S_6$ was examined through the frequency-dependent complex dielectric function.

The absorbance, $A(\omega)$, was evaluated from the absorption coefficient, $\alpha(\omega)$, using the following relation [52]:

$$A(\omega) = 1 - e^{-\alpha(\omega)\Delta z} \quad (1)$$

The absorption coefficient is calculated from the real and imaginary parts of the dielectric function as follows [52]:

$$\alpha(\omega) = \frac{\sqrt{2}\omega}{c}\left[\left(\varepsilon_1^2(\omega) + \varepsilon_2^2(\omega)\right)^{1/2} - \varepsilon_1(\omega)\right]^{1/2} \tag{2}$$

Here, $\omega$ denotes the angular frequency of the incident light, $c$ is the velocity of light in vacuum, and $\Delta z$ represents the unit-cell length along the out-of-plane direction. The terms $\varepsilon_1(\omega)$ and $\varepsilon_2(\omega)$ correspond to the real and imaginary components of the dielectric function, respectively. Among these two components, $\varepsilon_2(\omega)$ is directly associated with interband optical transitions and therefore provides important information about photon-induced electronic excitation.

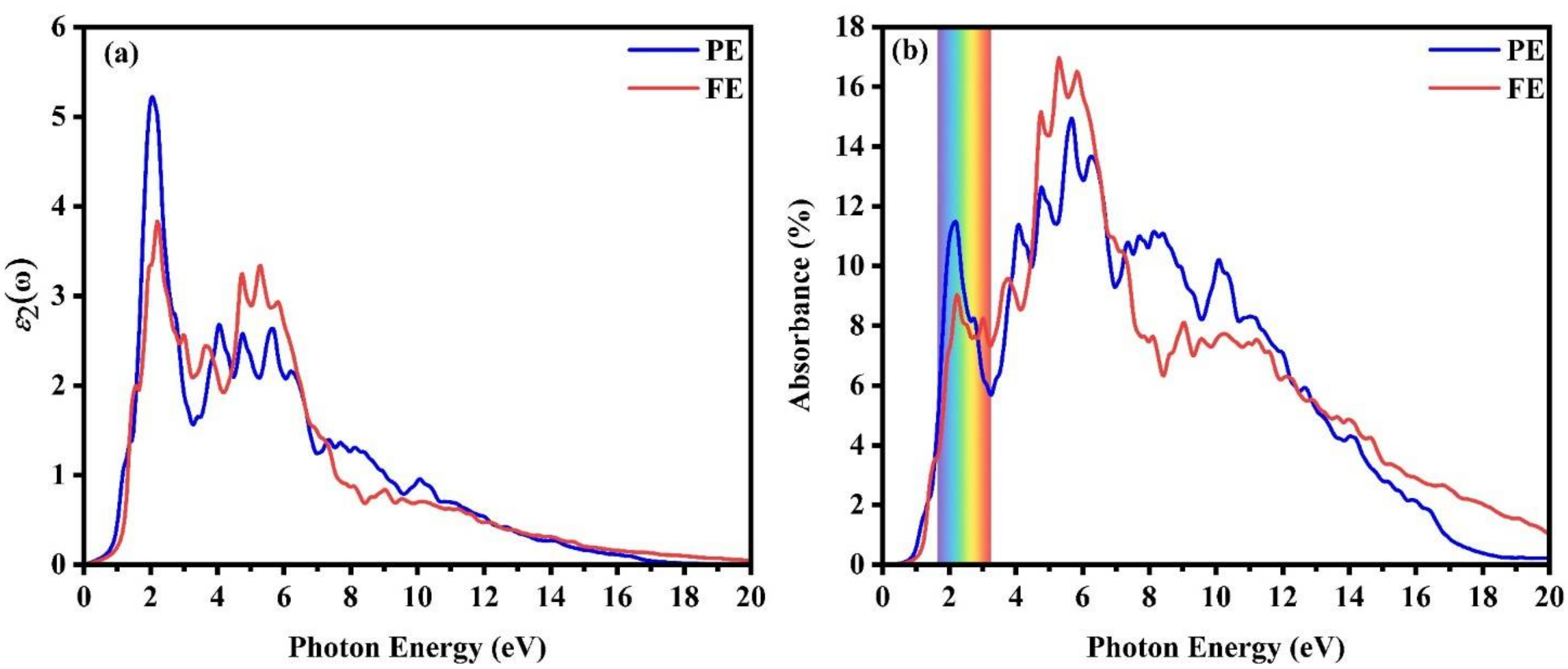

**Figure 6.** Photon energy dependences of (**a**) the imaginary parts of the dielectric function and (**b**) optical absorption spectra of PE and FE phases of $Zr_2Ge_2S_6$ monolayers.

As shown in **Figure 6(a),** the imaginary part of the dielectric function, $\varepsilon_2(\omega)$, demonstrates that both the FE and PE phases are optically active over a wide photon-energy range. In the low-energy region, the PE phase exhibits a pronounced peak near approximately 2 eV, where $\varepsilon_2(\omega)$ reaches about 5.2. The FE phase also shows a peak in the same energy region but its intensity is lower, with a maximum value of nearly 3.8. This indicates that the PE phase has a stronger optical transition probability in the visible-light region. After this initial intense response, both phases show several additional peaks between approximately 3 and 6 eV. In this intermediate energy range, the FE phase becomes more prominent than the PE phase, suggesting stronger electronic transitions in the ultraviolet region.

The absorbance spectra presented in **Figure 6(b)**, further confirm the phase dependent optical response of monolayer $Zr_2Ge_2S_6$. Within the visible-light region, highlighted approximately between 1.65 and 3.26 eV, both phases display noticeable light absorption. The PE phase reaches an absorbance of about 11-12% near the visible region, whereas the FE phase shows a comparatively lower absorbance of about 8-9% in the same energy window. This result suggests that the PE phase may be more favorable for harvesting visible photons.

At higher photon energies, the absorbance of both phases increases significantly. The FE phase reaches its maximum absorbance of nearly 16-17% around 5-6 eV, while the PE phase also exhibits strong ultraviolet absorption of approximately 14-15% in a similar energy range. Thus, although the PE phase shows better visible-light absorption, the FE phase becomes superior in the ultraviolet region. Beyond the main absorption peaks, especially above approximately 10 eV, the absorbance gradually decreases for both phases. However, the PE phase maintains a comparatively broader absorption profile over part of the high-energy region, whereas the FE phase shows a sharper reduction after its strongest ultraviolet peaks.

Overall, the calculated optical spectra indicate that monolayer $Zr_2Ge_2S_6$ possesses effective light-harvesting ability in both visible and ultraviolet regions. The PE phase exhibits stronger absorption in the visible-light range, while the FE phase provides more intense absorption in the ultraviolet range. This difference confirms that the optical properties of $Zr_2Ge_2S_6$ are strongly dependent on its polarization state. Therefore, the FE-PE phase transition can be considered an effective way to tune the photo response of monolayer $Zr_2Ge_2S_6$, which may be beneficial for photocatalytic and optoelectronic applications.

## 3.4 Mechanical properties and structural stability

Mechanical reliability is a central requirement for a 2D sheet because the lattice must tolerate tensile, compressive, and shear perturbations without losing structural integrity. Accordingly, the in-plane elastic response of the ferroelectric and paraelectric phases of monolayer $Zr_2Ge_2S_6$ was determined from the calculated stress-strain relations. Hexagonal symmetry reduces the in-plane linear elastic tensor to two independent stiffness coefficients, $C_{11}$ and $C_{12}$. The coefficient $C_{11}$ measures resistance to normal deformation along an in-plane direction, whereas $C_{12}$ describes the coupling between normal strains along two perpendicular directions. The shear coefficient is fixed by symmetry through $C_{66}$ and is therefore not an independent quantity. The in-plane stiffness tensor for both phases can consequently be written as follows:

$$\mathrm{C} = \begin{pmatrix} C_{11} & C_{12} & 0 \\ C_{12} & C_{11} & 0 \\ 0 & 0 & C_{66} \end{pmatrix} \tag{3}$$

The symmetry-dependent shear stiffness is therefore obtained from:

$$C_{66} = \frac{C_{11} - C_{12}}{2} \tag{4}$$

**Table 3.** The calculated elastic constants ($C_{ij}$ in $Nm^{-1}$), Young's modulus ($E$ in $Nm^{-1}$), shear modulus ($E$ in $Nm^{-1}$), bulk modulus ($B$ in $Nm^{-1}$), Poisson's ratio ($v$ in $Nm^{-1}$), bulk to shear ratio ($B/G$), the universal anisotropy index ($A_U$) and the stiffness eigenvalues ($\lambda_{1-3}$ in $Nm^{-1}$) of the PE and FE phases of $Zr_2Ge_2S_6$.

| Phase | $C_{11}$ | $C_{12}$ | $C_{66}$ | $E$ | $G$ | $B$ | $v$ | $B/G$ | $A_U$ | $\lambda_{1-3}$ |
|---|---|---|---|---|---|---|---|---|---|---|
| PE | 57.40 | 22.84 | 17.28 | 48.32 | 17.28 | 40.12 | 0.398 | 2.321 | 0.000 | 17.283, 34.567, 80.240 |
| FE | 63.57 | 21.47 | 21.05 | 56.32 | 21.05 | 42.52 | 0.338 | 2.020 | 0.000 | 21.052, 42.104, 85.040 |

For the FE phase, the calculated constants are $C_{11}$ = 63.57 Nm$^{-1}$, $C_{12}$ = 21.47 Nm$^{-1}$, and $C_{66}$ = 21.05 Nm$^{-1}$. The corresponding PE values are 57.40, 22.84, and 17.28 Nm$^{-1}$, respectively. Relative to the PE configuration, the FE phase has a 10.7% larger longitudinal stiffness and a 21.8% larger shear stiffness. Polarization strengthens the resistance of the $Zr_2Ge_2S_6$ lattice to both axial strain and in plane shape distortion. For a mechanically stable two-dimensional hexagonal lattice, the elastic constants must satisfy the Born conditions: $C_{11} > 0$ and $C_{11}^2 - C_{12}^2 > 0$ [53]; both phases meet these requirements.

The positive eigenvalues of the stiffness matrices provide an additional stability check. They are 21.052, 42.104, and 85.040 Nm$^{-1}$ for the FE phase and 17.283, 34.567, and 80.240 Nm$^{-1}$ for the PE phase. The findings that none of these eigenvalues is negative and elastic tensors are positive definite confirms stability against infinitesimal in-plane distortions.

The angular dependences of the in-plane Young's modulus, $E(\theta)$, and Poisson's ratio, $\nu(\theta)$, were evaluated from the elastic constants using the directional tensor formalism adopted for FE and PE two-dimensional monolayers in **Ref.** [35]:

$$E(\theta) = \frac{C_{11}C_{22} - C_{12}^2}{C_{22}\cos^4\theta + A\cos^2\theta\sin^2\theta + C_{11}\sin^4\theta} \tag{5}$$

$$\nu(\theta) = \frac{C_{12}\cos^4\theta - B\cos^2\theta\sin^2\theta + C_{12}\sin^4\theta}{C_{22}\cos^4\theta + A\cos^2\theta\sin^2\theta + C_{11}\sin^4\theta} \tag{6}$$

The auxiliary combinations A and B are defined as:

$$A = \frac{C_{11}C_{22} - C_{12}^2}{C_{66}} - 2C_{12} \tag{7}$$

$$B = C_{11} + C_{22} - \frac{C_{11}C_{22} - C_{12}^2}{C_{66}} \tag{8}$$

Here, $\theta$ denotes the angle between the applied uniaxial load and the chosen in-plane crystallographic axis. The coefficients $C_{11}$ and $C_{22}$ are the normal stiffnesses along two orthogonal directions, $C_{12}$ describes normal strain coupling and $C_{66}$ is the in-plane shear stiffness. The quantities $A$ and $B$ collect these coefficients into compact expressions for the directional mechanical response.

For a hexagonal monolayer, symmetry imposes the relations

$$C_{22} = C_{11}, \qquad C_{66} = \frac{C_{11} - C_{12}}{2} \tag{9}$$

Substitution of these symmetry relations into **Equation (5)-(8)** eliminates the angular terms and gives the reduced forms as follows:

$$E = \frac{C_{11}^2 - C_{12}^2}{C_{11}}, \qquad \nu = \frac{C_{12}}{C_{11}} \tag{10}$$

$$B = \frac{C_{11} + C_{12}}{2}, \qquad G = C_{66} = \frac{C_{11} - C_{12}}{2} \tag{11}$$

#### 3.4.1 Directional Young's modulus and Poisson's ratio

**Figure 7 and Figure 8** visualize the angular elastic response. A noncircular polar envelope indicates directional anisotropy, whereas a circular trace signifies in-plane rotational invariance following the interpretation used for FE and PE monolayers in **Ref.** [35]. The values plotted here are calculated specifically for $Zr_2Ge_2S_6$.

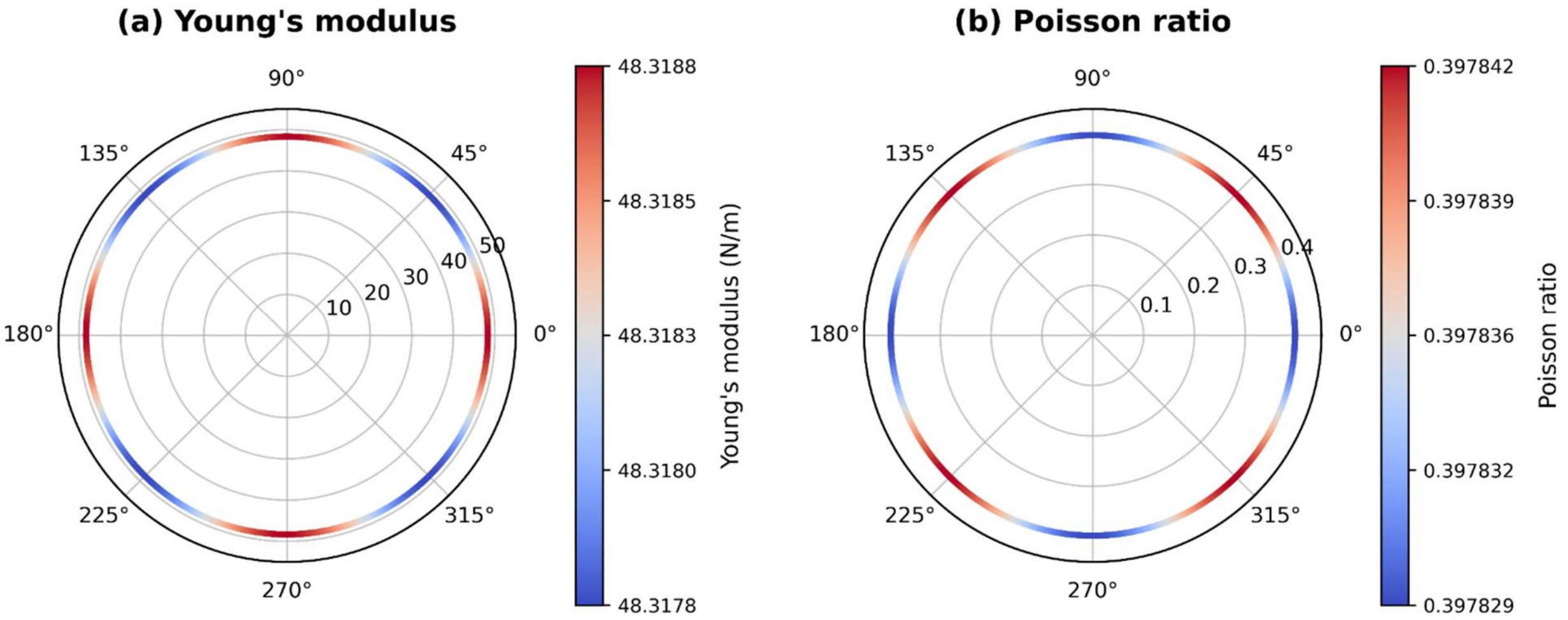

**Figure 7.** Angular dependence of (a) Young's modulus and (b) Poisson's ratio for the PE phase of monolayer $Zr_2Ge_2S_6$.

For the PE structure, $E(\theta)$ remains centered at approximately 48.318 $Nm^{-1}$ and $v(\theta)$ at approximately 0.398 throughout the full 0-360° rotation. The red to blue color changes spans only 48.3178 - 48.3188 $Nm^{-1}$ for Young's modulus and 0.397829 - 0.397842 for Poisson's ratio. These extremely narrow intervals are numerical scale fluctuations rather than evidence of physically meaningful anisotropy. The circular radial contours confirm that the PE phase has the same elastic response along every in-plane direction.

The FE curves are likewise circular, with Young's modulus fixed near 56.322 $Nm^{-1}$ and Poisson's ratio near 0.338. Compared with the PE phase, the FE phase has a 16.6% higher Young's modulus but an approximately 15.1% lower Poisson's ratio. Thus, ferroelectric ordering stiffens the monolayer against axial loading while reducing the magnitude of transverse contraction. Because the shape of the polar envelopes remains unchanged, the phase transition alters elastic strength without breaking in-plane isotropy. This observation agrees with the universal anisotropy index reported in **Table 3**.

The scalar moduli are consistent with the polar plots. The Young's modulus increases from 48.318 $Nm^{-1}$ in the PE phase to 56.322 $Nm^{-1}$ in the FE phase making the FE structure approximately 16.6% stiffer under in-plane uniaxial loading. The corresponding shear moduli are 17.283 and 21.052 $Nm^{-1}$, respectively. This shows that the PE lattice is more compliant under shape changing deformation. The bulk moduli are 40.120 $Nm^{-1}$ for PE and 42.520 $Nm^{-1}$ for FE. Thus, the phase transformation changes the resistance to uniform in-plane compression by only about 6.0%.

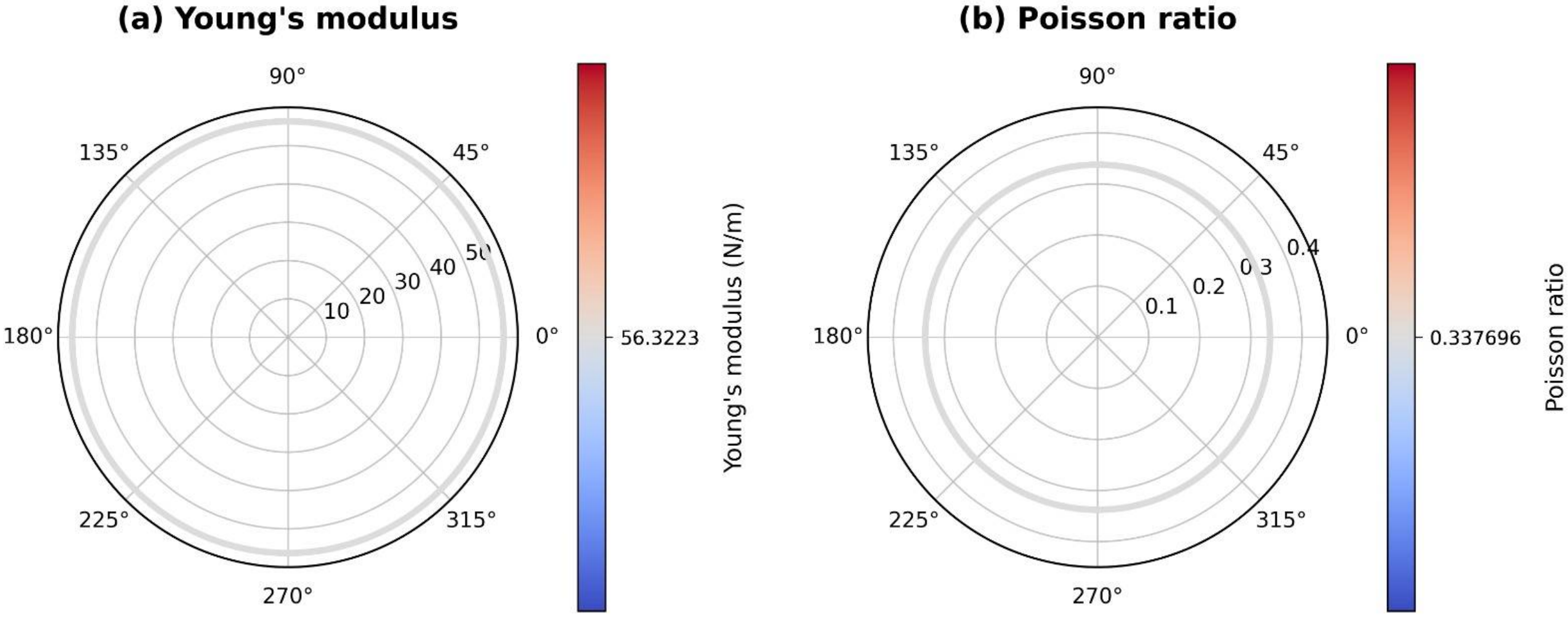

**Figure 8.** Angular dependence of (a) Young's modulus and (b) Poisson's ratio for the FE phase of monolayer $Zr_2Ge_2S_6$.

Both $Zr_2Ge_2S_6$ phases are mechanically softer than graphene (about 340 $Nm^{-1}$) [**54**], monolayer h-BN (about 275.8 $Nm^{-1}$) [**55**], and monolayer $MoS_2$ (about 114.5-128.9 $Nm^{-1}$) [**56**]. Their values lie within the broad direction-dependent interval reported for phosphorene (about 24-102 $Nm^{-1}$) [**57**]. This comparison suggests that FE and PE $Zr_2Ge_2S_6$ can accommodate appreciable elastic deformation under lower applied stress than several widely studied stiff two-dimensional sheets.

The Poisson's ratios are positive in both structures: 0.338 for FE and 0.398 for PE. Consequently, each phase contracts laterally under longitudinal tension and expands laterally under longitudinal compression. The larger PE value indicates stronger coupling between axial and transverse strains, whereas the smaller FE value shows that polarization weakens this transverse response. Neither phase is auxetic, because auxetic behavior requires a negative Poisson's ratio [**58**].

The bulk to shear ratios B/G are 2.020 for the FE phase and 2.321 for the PE phase. The larger PE ratio mainly originates from its reduced shear resistance relative to its areal compressibility. Since ductility criteria developed for three-dimensional bulk solids cannot be transferred directly to atomically thin materials, these ratios are interpreted only as comparative measures of the balance between compressional and shear response.

For each phase, the minimum and maximum directional values of Young's modulus, shear modulus, and Poisson's ratio coincide, giving an anisotropy ratio of 1.000. The Voigt and Reuss bounds likewise become identical:

$$\frac{B_{\mathrm{V}}}{B_{\mathrm{R}}} = 1.000, \qquad \frac{G_{\mathrm{V}}}{G_{\mathrm{R}}} = 1.000 \tag{12}$$

The universal anisotropy index is given by **Ref.** [**59**]:

$$A_{\mathrm{U}} = 0.000$$

The equality of the Voigt and Reuss limits, together with $A_U$ = 0 proves that both FE and PE $Zr_2Ge_2S_6$ are elastically isotropic within the monolayer plane. Their circular polar profiles are therefore not merely visual approximations. They are the direct consequence of hexagonal symmetry and the calculated elastic constants.

In summary, the Born stability conditions, positive stiffness eigenvalues, and isotropic polar responses collectively establish the mechanical stability of both phases of monolayer $Zr_2Ge_2S_6$. The FE phase is more resistant to longitudinal, shear, and uniform in-plane deformation, while the PE phase is softer and exhibits a stronger transverse strain response. The PE to FE transformation therefore changes the magnitude of the elastic response without generating directional anisotropy.

## 3.5 Band edge position

The photocatalytic potential of monolayer $Zr_2Ge_2S_6$ was assessed from the alignment of its band edges with the water redox potentials. In photocatalytic water splitting, the CBM must provide electrons with sufficient reducing power to drive HER, whereas the VBM must generate holes with enough oxidizing power to promote OER. Accordingly, the band edge positions of the PE and FE phases were referenced to the vacuum level, as shown in **Figure 9**, to evaluate their phase dependent redox capability.

The redox potentials of water were referenced to the vacuum level using the following relations [**60**]:

$$V_{\mathrm{H^+/H_2}} = E_{\mathrm{vacuum}} - 4.44 + \mathrm{pH} \times 0.059\ \mathrm{eV} \tag{13}$$

$$V_{\mathrm{O_2/H_2O}} = E_{\mathrm{vacuum}} - 5.67 + \mathrm{pH} \times 0.059\ \mathrm{eV} \tag{14}$$

The $H^+/H_2$ and $O_2/H_2O$ redox levels were referenced to vacuum at - 4.44 and -5.67 eV, respectively. The 1.23 eV difference between them defines the thermodynamic threshold for water splitting. Because these potentials are pH-dependent, both levels shift by 0.059 eV per pH unit according to the Nernst equation [**61**]. Therefore, the values shown in **Figure 9** correspond to the standard acidic limit, while increasing pH shifts both redox levels upward without altering their 1.23 eV separation.

In the PE phase, $Zr_2Ge_2S_6$ shows a symmetric electrostatic potential profile across the monolayer, as illustrated in **Figure 9(c)**. This symmetry arises from the centrosymmetric structure, where the absence of a permanent dipole prevents the formation of an internal polarization field. As a result, no appreciable potential difference develops between the upper and lower surfaces, giving $\Delta\Phi \approx 0$. Both sides of the PE monolayer therefore have nearly identical electrostatic environments and comparable band-edge alignments.

For the PE phase, the positive $\chi H_2$ value of 0.46 eV indicates that the CBM lies at a suitable reducing position relative to the $H^+/H_2$ redox level. Thus, photoexcited electrons have enough thermodynamic driving force to reduce protons or water derived hydrogen species into $H_2$. Because the PE structure is non-polar, its two surfaces are electrostatically equivalent, suggesting that HER can occur on either side of the monolayer. In contrast, $\chi O_2$ is negative, - 0.15 eV, showing that the VBM is not sufficiently oxidizing to drive water oxidation. The PE phase therefore favors HER but does not provide an adequate driving force for OER.

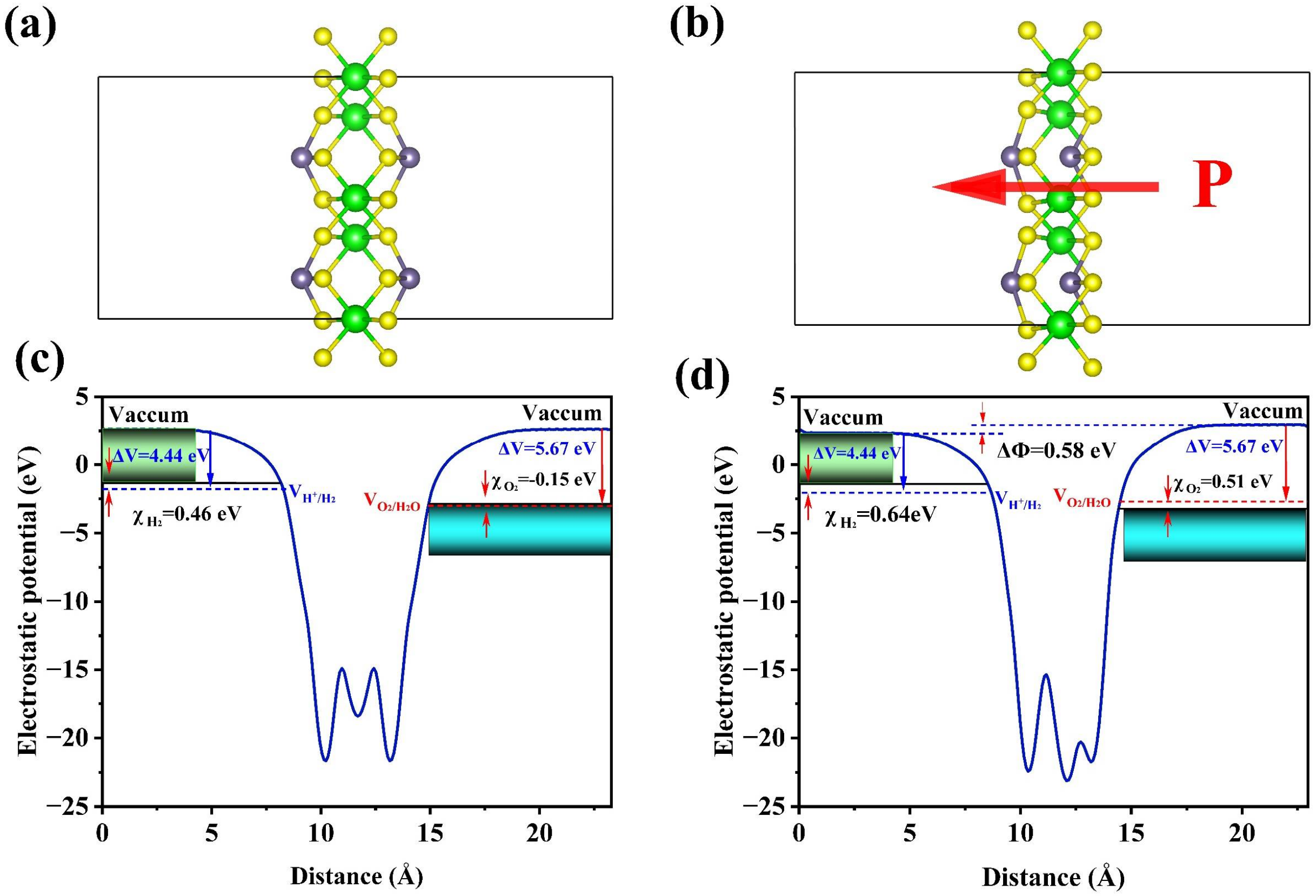

**Figure 9.** Potential profiles. (**a, b**) In-plane structural views of $Zr_2Ge_2S_6$ in the PE and FE phases. (**c, d**) Planar-averaged electrostatic potential profiles of the corresponding phases. The blue and red dashed lines represent the $H_2$ reduction and $H_2O$ oxidation potentials, while the green and cyan regions mark the CBM and VBM, respectively. $\chi(H_2)$ and $\chi(O_2)$ denote the HER and OER driving forces. $\Delta\Phi$ refers to the vacuum-level difference between the two surfaces, which reaches 0.58 eV in the FE phase and significantly strengthens the oxidation potential.

The FE phase shows a distinctly asymmetric electrostatic potential profile, arising from its out-of-plane ferroelectric polarization. As indicated by the polarization direction in **Figure 9(b)**, this built-in dipole creates an internal electric field across the monolayer. Consequently, the two surfaces no longer share the same vacuum level. The planar-averaged electrostatic potential in **Figure 9(d)** gives a surface potential difference of $\Delta\Phi$ = 0.58 eV, confirming the polarization-induced electrostatic imbalance. This built-in potential shifts the relative band edge positions at the opposite surfaces and enables surface dependent redox behavior.

In the FE phase, $\chi H_2$ increases to 0.64 eV, indicating a stronger reducing ability than in the PE phase. However, this activity is not expected to be uniform across the monolayer, because ferroelectric polarization makes the two surfaces electrostatically inequivalent. Thus, HER is likely to occur preferentially on the reduction-favorable surface. More notably, the OER driving force changes from $\chi O_2$ = -0.15 eV in the PE phase to $\chi O_2$ = 0.51 eV in the FE phase. This positive value shows that the VBM becomes favorably aligned for water oxidation. Therefore, the ferroelectric polarization not only separates the surface potentials but also activates the OER pathway in $Zr_2Ge_2S_6$.

Therefore, the two phases demonstrate that the photocatalytic function of $Zr_2Ge_2S_6$ is strongly phase dependent. The PE phase is more suitable for HER because its CBM lies at a favorable position for $H_2$ generation while its VBM is insufficient for OER. On the other hand, the FE phase generates an internal electric field that separates the two surface potentials and improves the oxidative capability of the material. This polarization induced band bending enhances OER activity and produces surface selective redox behavior.

## 3.6 Tunable overall water splitting reaction

To evaluate the photocatalytic activity of monolayer $Zr_2Ge_2S_6$ beyond band edge alignment, the Gibbs free-energy profiles of OER and HER were calculated under different phase and pH conditions (**Figure 10**). While band positions indicate whether water splitting is thermodynamically viable, the reaction feasibility is ultimately controlled by the energy barriers of the elementary steps. Thus, the OER and HER pathways were examined to clarify the catalytic roles of the FE and PE phases.

As shown in **Figure 10(a)**, OER on the FE phase follows the four-step pathway $H_2O \rightarrow {*OH} \rightarrow {*O} \rightarrow {*OOH} \rightarrow O_2$. At pH = 0 and U = 0 V, the reaction is energetically uphill, with *OOH formation as the rate-determining step and a barrier of 2.04 eV. When the photogenerated hole potential is applied, U=1.74 V, this barrier decreases sharply to 0.29 eV. This indicates that illumination provides the oxidative driving force needed to make OER much more favorable on the FE phase.

The OER activity of the FE phase also depends strongly on pH. At pH = 7 and U = 0 V, the rate-determining barrier decreases from 2.04 to 1.62 eV, showing that neutral conditions are more favorable for water oxidation than acidic conditions. With the photogenerated hole potential of U = 2.16 V, the whole OER pathway becomes downhill. This suggests that illuminated FE- $Zr_2Ge_2S_6$ can drive OER efficiently under neutral conditions. The improvement mainly arises from the pH-induced shift of the redox levels, which enhances the effective oxidative driving force of the holes.

HER was evaluated at pH = 0 for both phases. In the FE phase, **Figure 10(b)**, proton adsorption to form *H requires 0.94 eV at U = 0 V, indicating sluggish HER. Under illumination, the electron driving force of U = 0.64 V lowers this barrier to 0.30 eV and makes $H_2$ formation downhill. Thus, the FE phase can support HER under photoexcitation, although a small adsorption barrier remains.

The PE phase shows more favorable HER behavior. As shown in **Figure 10(c)**, *H formation requires only 0.36 eV at U = 0 V, much lower than in the FE phase. With the photogenerated electron potential of U = 0.46 V, this step becomes nearly thermoneutral, followed by downhill $H_2$ formation. These results indicate that the PE phase is the preferred reduction active state for HER in $Zr_2Ge_2S_6$.

The OER and HER free-energy profiles reveal a clear phase-selective catalytic behavior in monolayer $Zr_2Ge_2S_6$. The FE phase favors water oxidation, especially under illumination at neutral pH, whereas the PE phase provides a lower-barrier pathway for hydrogen evolution. This division of activity suggests a tunable water-splitting mechanism, where the PE phase acts as the HER-active state for $H_2$ production and the FE phase serves as the OER-active state for $O_2$ generation.

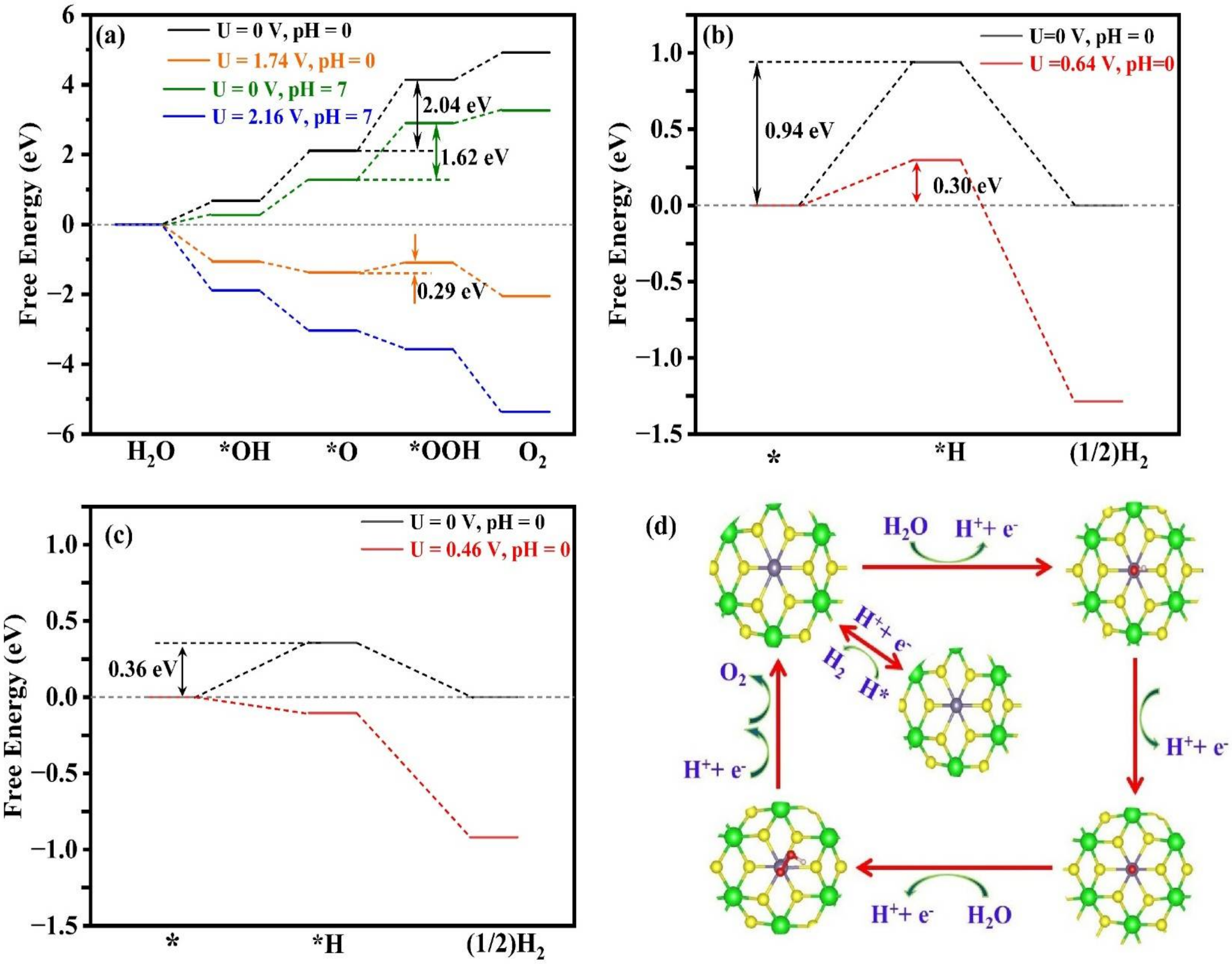

**Figure 10.** Reaction driving force: (a) Gibbs free energy of OER of FE phase at pH = 0 and pH =7 (**b**) HER at pH = 0 (**c**) HER for the PE phase at pH = 0 (**d**) Photocatalytic path of OER and HER.

**Figure 10(d)** summarizes the proposed switchable water-splitting pathway of monolayer $Zr_2Ge_2S_6$. The same material can selectively drive different half-reactions by changing its phase: the PE state favors HER, while the FE state promotes OER. This built-in phase selectivity offers a simple route to tune overall water splitting without requiring separate cocatalysts or complex heterostructures. Thus, the reaction pathway is mainly controlled by the ferroelectric phase, photogenerated carrier driving force and solution pH.

Overall, the free-energy results show that water splitting in $Zr_2Ge_2S_6$ can be tuned through the combined effects of phase state and pH. The FE phase is more favorable for OER and can drive downhill water oxidation under illuminated neutral conditions. In contrast, the PE phase provides the more efficient HER pathway, with nearly barrierless hydrogen adsorption under photoexcitation. These results identify $Zr_2Ge_2S_6$ as a phase-switchable photocatalyst for controllable overall water splitting.

## 3.7 Energy conversion efficiency

The solar-to-hydrogen efficiency of monolayer $Zr_2Ge_2S_6$ shows a clear dependence on the PE and FE phase states. Although the PE phase has a higher light absorption efficiency of 48.98%, its carrier utilization efficiency is limited to 15.75%, resulting in an STH efficiency of 7.71%. This suggests that strong photon absorption alone is not sufficient for efficient water splitting, as the photogenerated electrons and holes must also be effectively separated and used in the surface redox reactions.

**Table 4.** The calculated photocatalytic parameters of $Zr_2Ge_2S_6$ for the PE and FE phases, including light absorption efficiency $\boldsymbol{\eta}_{\mathbf{abs}}$, carrier utilization efficiency $\boldsymbol{\eta}_{\mathbf{cu}}$, STH efficiency $\boldsymbol{\eta}_{\mathbf{STH}}$, and corrected STH efficiency $\boldsymbol{\eta}'_{\mathbf{STH}}$. All energy values are in eV and all efficiencies are in %.

| Phase | $\eta_{\mathrm{abs}}$ | $\eta_{\mathrm{cu}}$ | $\eta_{\mathrm{STH}}$ | $\eta'_{\mathrm{STH}}$ |
|---|---|---|---|---|
| PE | 48.98 | 15.75 | 7.71 | ------ |
| FE | 36.06 | 42.46 | 15.31 | 14.13 |

In the FE phase, the light absorption efficiency decreases to 36.06%, but the carrier utilization efficiency rises sharply to 42.46% due to the polarization induced internal electric field and the vacuum-level difference of 0.58 eV. This enhanced carrier utilization increases the STH efficiency to 15.31%, with a corrected value of 14.13%. Therefore, FE-$Zr_2Ge_2S_6$ exhibits superior solar-to-hydrogen performance mainly because ferroelectric polarization improves charge separation and strengthens the overall redox utilization of photogenerated carriers.

# 4 Conclusion

This work identifies monolayer $Zr_2Ge_2S_6$ as a promising phase-switchable photocatalyst in which a paraelectric-ferroelectric transition modulates the electronic structure, interfacial electrostatics, and water-splitting reactivity. The calculations indicate that both PE and FE phases are structurally, dynamically, and thermally stable. The phase transition introduces a meaningful change in band gap, band edge alignment, and charge distribution. According to our calculations, ferroelectric switching can regulate the redox capability and driving force of photogenerated carriers in monolayer $Zr_2Ge_2S_6$. In particular, the PE phase exhibits stronger reduction behavior because its photogenerated electrons possess a greater driving force. In contrast, the FE phase gains a polarization induced oxidation advantage due to the enhanced driving force of photogenerated holes. These differences make the two phases functionally distinct for HER and OER related reactions.

The significance of the study lies in how ferroelectricity reorganizes the photocatalytic landscape. In the FE phase, the built-in polarization generates an internal electric field and a calculated vacuum level difference of 0.58 eV which produces asymmetric band edge states and more effective electron-hole separation. This behavior is consistent with the broader understanding that ferroelectric polarization can suppress recombination and promote surface selective redox chemistry in photocatalysis. Within our dataset, that mechanism is reflected in the enhanced oxidation driving force, improved carrier utilization, and the rise in calculated STH efficiency from 7.71% in the PE phase to 15.31% in the FE phase, even though visible light absorption is stronger in the PE phase.

Future work should therefore focus on linking the predicted phase-selective HER/OER behavior to operando conditions, including pH, illumination intensity, strain, and electrode/electrolyte environment. So that $Zr_2Ge_2S_6$ can be assessed not only as a conceptually elegant 2D ferroelectric photocatalyst, but also as a practically addressable platform for controllable solar fuel generation.

## Data availability

The data sets generated and/or analyzed in this study are available from the corresponding author on reasonable request.

## Declaration of interest

The authors declare no competing interests.

## CRediT authorship contribution statement

**Jubair Hossan Abir:** Conceptualization, Software, Methodology, Formal analysis, Data curation, Visualization, Writing-original draft, review & editing; **Tauhidur Rahman:** Conceptualization, Software, Methodology, Formal analysis, Data curation, Visualization, Writing-draft, review & editing; **Tanvir Khan:** Methodology, Formal analysis, Data curation, Visualization, Writing-draft, review & editing; **Sarker Sukanta Babu Pallab:** Visualization, Writing- draft, review & editing; **Raihana Shams Islam:** Conceptualization, Supervision, Validation, Administration, Writing- Reviewing and Editing; **Saleh Hasan Naqib:** Conceptualization, Supervision, Validation, Administration, Writing- Reviewing and Editing.

## Declaration of Generative AI and AI-assisted technologies in the writing process

During the preparation of this work, the authors used ChatGPT (OpenAI) in order to improve the language, grammar, readability, and organization of the manuscript. After using this tool, the authors reviewed and edited the content as needed and take full responsibility for the content of the published article.